# Experimental observation of ferrielectricity in multiferroic DyMn$_2$O$_5$: revisiting the electric polarization

Z. Y. Zhao[1], M. F. Liu[1], X. Li[1], L. Lin[1], Z. B. Yan[1], S. Dong[2], and J.–M. Liu[a]

[1]*Laboratory of Solid State Microstructures, Nanjing University, Nanjing 210093, China*
[2]*Department of Physics, Southeast University, Nanjing 210189, China*

**[Abstract]** The electric polarization and its magnetic origins in multiferroic RMn$_2$O$_5$, where R is rare-earth ion, are still issues under debate. In this work, the temperature-dependent electric polarization of DyMn$_2$O$_5$, the most attractive member of this RMn$_2$O$_5$ family, is investigated using the pyroelectric current method upon varying endpoint temperature of the electric cooling, plus the positive-up-negative-down (PUND) technique. It is revealed that DyMn$_2$O$_5$ at low temperature does exhibit the unusual ferrielectricity rather than ferroelectricity, characterized by two interactive and anti-parallel ferroelectric sublattices which show different temperature-dependences. The two ferroelectric sublattices are believed to be generated from the symmetric exchange-striction mechanisms associated with the Mn-Mn spin interactions and Dy-Mn spin interactions, respectively. The path-dependent electric polarization reflects the first-order magnetic transitions in the low temperature regime. The magnetoelectric effect is mainly attributed to the Dy spin order which is sensitive to magnetic field. The present experiments may be helpful for clarifying the puzzling issues on the multiferroicity in DyMn$_2$O$_5$ and probably other RMn$_2$O$_5$ multiferroics.

---

[a] Correspondent, E-mail: liujm@nju.edu.cn

# I. Introduction

Multiferroics have been intensively investigated for ten years since the pioneer works on $BiFeO_3$ [1] and $TbMnO_3$ in 2003 [2]. In particular, the discovery of so-called type-II multiferroics has intrigued comprehensive understanding of multiferroicity [2-4]. In these materials, electric polarization $P$ is believed to be intrinsically correlated with particular magnetic ordering or re-ordering below certain temperatures and thus the cross-coupling between ferroelectricity and magnetism is usually significant, allowing possible magnetic control of electric polarization or/and electric control of magnetism [5-11]. To dates, what keeps much of the research interest alive is the possibility of discovering underlying microscopic physics which is substantially different from our earlier knowledge on multiferroicity and even general principles for guiding design and synthesis of multiferroic materials of promising practical applications [5, 12-15], while it is noted that so far most discovered type-II multiferroics either have low transition temperatures or only offer small polarization and weak ferromagnetism.

While conventional ferroelectrics accommodate electric polarization via the symmetry-breaking transitions from high symmetric paraelectrics (PE) phase [16], for these multiferroics the primary order parameter is magnetic rather than structural. Two major magnetic mechanisms for the ferroelectricity have been proposed. One is the asymmetric exchange striction scenario (category I), in which the inverse Dzyaloshinskii-Moriya (DM) interaction associated with the non-collinear spin order (helical or cycloidal spin structure) drives the structural symmetry-breaking [17, 18]. The typical materials echoing this scenario include rare-earth manganites $RMnO_3$ (R=Gd, Tb, Dy, Ho etc) [19, 20], barium ferrites [21], $LiCuO_2$ [22], $Ni_3V_2O_8$ [23], and many others [24]. The other is the symmetric exchange striction scenario (category II), in which specific collinear spin orders, such as E-type antiferromagnetic (AFM) order in some orthorhombic $RMnO_3$ ($HoMnO_3$, $YMnO_3$) [25, 26] and ↑↑↓↓ order in $Ca_2CoMnO_6$ etc [27], favors the structural symmetry-breaking. The asymmetric exchange striction usually allows remarkable polarization response to magnetic field while the polarization itself is small. For the materials in the category II scenario, the polarization can be large but the magnetic control of polarization is much weaker. Besides the two mechanisms, it is argued that $Ba_2CoGe_2O_7$ [28] and $CuFeO_2$ [29] accommodate the *p-d* hybridization as the origin of ferroelectricity, and the ferroaxial model was proposed to track the electric polarization in $CaMn_7O_{12}$ [30, 31].

Interestingly, another class of multiferroics, in which the symmetric exchange striction is believed to play major role while the role of the asymmetric exchange striction is non-negligible too, is rare-earth manganites $RMn_2O_5$ family [32]. In comparison with the aforementioned materials in the two categories, $RMn_2O_5$ shows more complicated lattice and spin structures, accommodates multifold interaction competitions, larger electric polarization, and also more significant magnetoelectric responses [5, 16, 33]. In this sense, some researchers advised this $RMn_2O_5$ family as the third class of multiferroics other than the above mentioned two classes [16], and thus substantial attentions have been paid to this family [34].

Nevertheless, partially due to the complexity of lattice distortion and magnetic structure, the multiferroic transitions and underlying mechanisms in $RMn_2O_5$ are still much under debates [8, 9]. The $RMn_2O_5$ family has similar structural ingredients [16]. The Mn ions are partitioned into $Mn^{3+}$ and $Mn^{4+}$, which are coordinated respectively in square pyramid Mn-O units and octahedral Mn-O units. The projected lattice on the *ab*-plane is shown in Fig.1, where the oxygen ions all occupying the corners of the pyramids and octahedra are labeled with small dots. On the *ab*-plane, the octahedra and pyramids are corner-sharing by either the pyramid base or pyramid apex, and the adjacent pyramids are connected with their base. Along the *c*-axis, the octahedra sharing edges constitute linear chains. Each $Mn^{3+}$ ion is located in between two $Mn^{4+}$ ions, and the $R^{3+}$ ions are located on the alternative layer between two $Mn^{4+}$ ions. Therefore, $RMn_2O_5$ can be written as $R^{3+}Mn^{3+}Mn^{4+}5O^{2-}$, and the cation alignment sequence along the *c*-axis is zigzag-like $Mn^{3+}$-$Mn^{4+}$-$R^{3+}$-$Mn^{4+}$-$Mn^{3+}$-$R^{3+}\cdots$.

Due to this complicated structure, the Mn spin interactions become multi-folded, characterized by three dominant interactions ($J_3$, $J_4$, and $J_5$, as shown in Fig.1) plus even longer-range interactions [16, 35]. Therefore, spin frustration becomes a major feature of the magnetic transitions, leading to consecutive commensurate antiferromagnetic (C-AFM) and incommensurate AFM (IC-AFM) ordering sequence [36, 37]. Furthermore, the 3*d*-4*f* interactions in $RMn_2O_5$ with magnetic R ion can't be neglected and the strong R-Mn coupling allows even more fascinating spin structure evolution upon variations in temperature and external field. For example, the Mn and Tb magnetic sublattices in $TbMn_2O_5$ can be fully decoupled while it is not the case for $HoMn_2O_5$ and $DyMn_2O_5$ [35, 38, 39]. These facts reflect the substantial role of the R-Mn interactions in addition to the Mn-Mn interactions. As a consequence, the magnetic transitions become particularly material-dependent. Upon nonmagnetic or magnetic R ion and its size variation, the transition sequence can be different [40]. In correspondence, the ferroelectric (FE) transitions associated with these magnetic transitions are strongly material-dependent in addition to the complexity [8, 13]. It is also due to these complexities that the multiferroic behaviors, particularly the ferroelectricity and its magnetic origins, among many other issues are still far from well understood.

Along this line, we take $DyMn_2O_5$ as a representative case for illustration, noting that it is a member deserving for specific interests in past few years [8, 9, 34]. It is generally believed that the paramagnetic phase above temperature $T$~43K transits into an IC-AFM phase, followed by a C-AFM phase below $T_{N1}$=~40K, and then by the coexistence of an IC-AFM phase and a C-AFM phase below $T_{N2}$~28K. This coexistence is again replaced by two coexisting IC-AFM phases below $T_{N3}$~20K. At $T<T_{Dy}$~8K, the $Dy^{3+}$ spins order independently. The structural and interaction origins for these magnetic transitions were discussed extensively, while no full consistency has been reached [8, 9, 41]. Recently, the noncollinear Mn spin order with helical or cycloidal geometry in $DyMn_2O_5$ was reported [34].

In comparison with the magnetic structures and transitions, our understanding of the ferroelectricity is still in the earlier stage, and so far reported data are somehow controversial. While it is believed that the C-AFM phase is ferroelectric and the IC-AFM phase ICP is not, the measured results are not always consistent with this prediction [8, 9]. Basically, the measured $P$ most likely aligns along the $b$-axis although any component along other directions has not yet been excluded. However, the measured $P(T)$ data in the experiment by Hur *et al* using conventional pyroelectric current method show that the $P$ initiates below $T_{N1}$ ($T_{FE1}$) and changes its sign from negative value to positive one at some $T$ lower than $T_{N2}$~27K. A ferrielectric (FI) state with two antiparallel polarization components below $T_{N2}$ was then proposed [8]. In the experiment by Higashiyama *et al* using a different probing method [9], it was observed that the measured $P(T)$ experiences several transitions which seem to be one-to-one corresponding with the magnetic transitions, and the system becomes non-ferroelectric below $T_{Dy}$~8K (the so-called X-phase). The FI state as proposed by Hur *et al* [8] seems to be no longer aware in subsequent works on DyMn$_2$O$_5$. Besides, the magnetic origins for these transitions were discussed in details but no conclusive scenario on the polarization generation is available [16].

Conventionally available data on the magnetoelectric effect in RMn$_2$O$_5$ family are also scattered. Several general features can be outlined, although exceptional cases are available. First, remarkable magnetoelectric response, characterized by magnetic field induced tremendous changes in *P* and dielectric constant $\varepsilon$, is observed for those materials with magnetic R ion [8, 9, 12, 36, 37], while only weak or medium magnetoelectric response for those with medium or non-magnetic R ion (e.g. non-magnetic Y and Bi, relatively weak magnetic Er, Tb, and Yb) [8, 42]. Second, the magnetoelectric effect appears in the low-*T* regime where the R-Mn coupling is strong enough to be no longer negligible in determining the magnetic structure. These features allow one to argue that the magnetoelectric effect may be mainly attributed to the R-Mn spin coupling, while the Mn spin structure can be robust against magnetic field up to ~10T [8, 9, 37]. Nevertheless, no comprehensive understanding of these features is available to us.

The above inconsistencies and the insufficient data on the ferroelectricity and magnetoelectric effect leave substantial uncertainties in order to understand the ferroelectric behavior and its correlations with magnetic structure as well as the underlying mechanisms. This arises critical appealing for revisiting the electric polarization and its response to magnetic field in RMn$_2$O$_5$ (here DyMn$_2$O$_5$) as a function of *T*, so that these issues can be better clarified. On the basis that DyMn$_2$O$_5$ facilitates strong Dy-Mn coupling in addition to the dominant Mn-Mn interactions [38], one has reasons to expect a ferrielectric (FI) state with more than one polarization component. With no doubt, convincing evidences conforming this ferrielectric state become primarily critical for understanding the multiferroicity of DyMn$_2$O$_5$ and more generally the RMn$_2$O$_5$ family.

In proceeding, one first needs to check conventionally employed methods for measuring the electric polarization and develop an alternative method to provide details of the electric polarization in the ground state without external field bias, so that a comprehensive scenario on the ferroelectricity and its dependence on magnetic structure can be reached. This is the major motivation of the present work. We shall start from a discussion on the two methods employed for probing the electric polarization, and then present our data obtained by a modified pyroelectric current method, to be additionally complimented by the positive-up-negative-down (PUND) method. It will be suggested that $DyMn_2O_5$ is a ferrielectric composed of two anti-parallel ferroelectric sublattices, rather than a ferroelectric. The electric polarizations of the two sublattices have different origins, with one from the Mn-Mn symmetric exchange striction and the other from the Dy-Mn symmetric exchange striction. It is noted that the magnetoelectric effect can be reasonably explained by this ferrielectric scenario.

The remaining of this article is organized as the follow. In Sec.II we discuss several issues on the methodology for the electric polarization measurements, and propose the modified pyroelectric current method. The sample preparation and characterization details will be given in Sec.III. The main results and discussion will be presented in Sec.IV. A simple model for the ferrielectricity will be proposed in Sec.V, with a reasonable interpretation of the magnetoelectric data, followed by a brief conclusion in Sec.VI.

**II. Issues on methodology for measuring electric polarization**

*A. Methods for measuring polarization P*

So far available data on electric polarization $P$ as a function of $T$ and magnetic field $H$ were obtained by means of three different methods [8, 9].

The first is the conventional pyroelectric current (Pyro) method, which has been extensively used in measuring $P(T, H)$ for multiferroics with extremely small polarization (as small as $\sim 1.0 \mu C/m^2$) and low transition temperature [27]. The Pyro method is much more sensitive than the Saywer-Tower method or virtual-ground method conventionally used for the $P$-$E$ hysteresis of normal ferroelectrics. A schematic illustration of the Pyro method is given in Fig.2(a). The sample is submitted to electric poling under a field $E_{pole}$ during the cooling run until $T=T_{end}$, and then short-circuited for sufficient time at $T_{end}$. $T_{end}$ should be as low as possible and for typical case $T_{end}=2K$. The released current $I_{tot}$ from the sample is probed with electric bias $E_m=0$ during the subsequent warming run from $T_{end}$ up to an assigned temperature $T_0$. Certainly, $T_0$ should be much higher than the highest FE transitions.

If the released current $I_{tot}$ only contains the pyroelectric current $I_{pyro}$ without other contribution, $I_{tot}=I_{pyro}=0$ will be observed. The $I_{pyro}(T)$ is integrated from $T_0$ down to $T_{end}$, generating electric polarization $P_{pyro}$ as a function of $T$. The Pyro method is applicable only if $P$ at $T_{end}$ is nonzero, otherwise the electric poling down to $T_{end}$ is completely ineffective even if $P$ is nonzero at any $T$ other than $T_{end}$.

An often questioned issue for this method is that the current signals $I_{tot}$ possibly include contributions other than $I_{pyro}$, such as poling induced trapped charges or/and thermally stimulated current. A well-recognized way to exclude these contributions is to perform the measurement at several runs with different warming rates. In case of no shift between the measured $I_{tot}(T)$ curves along the $T$-axis, $I_{tot}=I_{pyro}$ is recognized. This method was used in the earliest experiment on polarization of $RMn_2O_5$ [8]. Surely, due to the uncertainties associated with the apparatus, a shift of the curve as small as ~1.0K along the $T$-axis is possible, which is the error of the measuring apparatus.

The second is the high precision *P-E* loop (P-E) method, which is applicable for ferroelectrics with relatively large *P*. Its successful application to $DyMn_2O_5$ was reported, and well-defined *P-E* loops were obtained above $T_{Dy}$~8K [9]. However, no identifiable loop can be observed below $T_{Dy}$, by which one infers that the low-*T* phase is non-ferroelectric (the X-phase). For $DyMn_2O_5$, it was claimed that the direction of *P* is neither uniquely nor controllably fixed if the Pyro method is used [9], and this failure was thought to be related to the ineffective poling process if $T_{end}<T_{Dy}=8K$. Surely, this P-E method may be questioned if *P* is relatively small in terms of the *P(T)* dependence.

Based on the above discussion, Higashiyama *et al* [9] developed the third method and we may name it as the Pole method, as schematically drawn in Fig.2(b). Instead of separating the poling step and probing step in the Pyro method, here the poling and probing are carried out simultaneously. The sample under a poling field $E_{pole}$ is gradually cooled down to $T_{end}$ from a high *T* during which the total current $I_{tot}$ across the sample is probed. This field is supposed to be small sufficiently so that the field induced leaky current $I_{leaky}$ is much smaller than $I_{pyro}$ over the whole *T*-range covered experimentally. A proper fitting procedure may allow an exclusion of $I_{leaky}$, leaving $I_{pyro}(T)$ and thus $P(T)$ to be evaluated, respectively. Clearly, an immediate question regarding this Pole method is the validity of $I_{leaky}<<I_{pyro}$, which may not be always true even for DyMn$_2$O$_5$, since a small $E_{pole}$ implies an incomplete poling of the sample and thus the evaluated *P* may not be the spontaneous polarization. Furthermore, a separation of $I_{pyro}$ from $I_{tot}$ is anyway a matter if dependence relations $I_{leaky}(T)$ and $I_{pyro}(T)$ are unknown, while relation $I_{leaky}(T)$ is not easily accessible. In fact, Higashiyama *et al* discussed the consequences of these issues in their work [9]. The major issue here for DyMn$_2$O$_5$ is that this method may not identify correctly the ferrielectric state if DyMn$_2$O$_5$ would be a ferrielectric.

*B. Modifying the pyroelectric current method*

In Fig.3(a)-(c) are plotted the measured *P*(*T*) curves for DyMn$_2$O$_5$ single crystals by the above mentioned three methods (Pyro, P-E, Pole). The boundaries between different FE phases, as given by Higashiyama *et al* [9] are marked and labeled above Fig.3(a). The measured data on polycrystalline samples by the PUND method (to be described below) are inserted in Fig.3(d) for comparison too.

While no discussion on details of these measured *P*(*T*) data is given here, we only look at the correlations between the magnetic transition points and FE transition points. The *P*(*T*) curves obtained by the Pyro method (Fig.3(a)) and the PUND method (Fig.3(d)) show anomalies roughly at $T_{N1}$, $T_{N2}$, $T_{N3}$, and $T_{Dy}$, respectively, those curves obtained by the P-E method (Fig.3(b)) and the Pole method (Fig.3(c)) show no anomalies at $T_{N2}$ and $T_{N3}$. In particular, both the Pyro method and PUND method revealed that the X-phase is ferroelectric with considerable polarization at $T<T_{Dy}$. In fact, the measured $P_{P-E}(T)$ and $P_{pyro}(T)$ curves show remarkable differences. An anomaly at *T*~13K, $T_{FE3}$ here, was observed in the measurements by the P-E, Pole, and PUND methods, but not by the Pyro method, noting that only the Pyro method facilitates zero electric bias during the measurement.

The above comparison stimulates us to revisit the Pyro method. We then modify this method (i.e. the mPyro method). Instead of cooling the sample down to the lowest *T*, e.g. $T_{end}$~2K<<$T_{Dy}$, we take $T_{end}$ as a variable upon requirement. Given that the suggested X-phase is non-ferroelectric at $T<T_{Dy}$, we choose a series of $T_{end}$ and perform identical measurements. This mPyro method offers several advantages. Besides the aforementioned one, one may also avoid possible influence of the magnetic transitions below $T_{end}$ on the multiferroic behaviors above $T_{end}$, which is believed to be significant for a number of type-II multiferroics such as MnWO$_4$ [43] and RMnO$_3$ (R=Gd, Dy, Ho) [44]. The possibly existing issues in the P-E and Pole methods, such as leaky current contribution and uncertainties, may be resolved too.

It will be shown that the data obtained by this mPyro method are qualitatively similar to those by the Pyro and PUND methods but different from the other two. By step-by-step varying $T_{end}$ up to $T_{N1}$ from the lowest $T$ reachable, one is able to evaluate the electric polarization in various magnetic phases.

### III. Experimental details
#### A. Samples preparation & structural characterization

In this work, we focus on polycrystalline DyMn$_2$O$_5$ rather than single crystals based on two considerations. First, due to the possible uncertainties with the polarization measurement, whether the polarization only has the *b*-axis component or not remains to be clarified. This clarification is challenging. It is a proper choice to start from polycrystalline samples so that the orientation dependence can be avoided for simplification. Second, we have paid our attentions to the effect of various substitutions (not shown here) and it is much easier to access doped polycrystalline samples than to access single crystal ones.

The samples were prepared by standard solid state sintering. Stoichiometric quantities of Dy$_2$O$_3$(99.99%) and Mn$_2$O$_3$(99%) were thoroughly mixed, compressed into pellets, and sintered at 1200$^o$C for 24 h in an oxygen atmosphere with several cycles of intermediate grindings. For every sintering cycle, the samples were cooled at 100$^o$C per hour from the sintering temperature down to room temperature [35]. The as-prepared samples were cut into various shapes for subsequent microstructural and property characterizations. The sample crystallinity was checked using X-ray diffraction (XRD) with Cu K$\alpha$ radiation at room temperature and the obtained $\theta$-$2\theta$ spectrum is presented in Fig.4. The reflections can be well indexed by the lattice symmetry *Pbam*, as also confirmed with the refined data using the Rietveld analysis. The evaluated lattice constants are $a$=0.7298(4) nm, $b$=0.8510(5) nm, and $c$=0.5681(8) nm with factor $R_{WP}$=6.41%. These data are consistent with earlier reported values [38].

### B. Measurement of magnetic and electric properties

Extensive measurements on the specific heat, magnetization and magnetic susceptibility, dielectric and ferroelectric properties of the samples were carried out. The magnetization $M$ and magnetic susceptibility $\chi$ were measured using the Quantum Design Superconducting Quantum Interference Device (SQUID) in the zero-field cooled (ZFC) mode and field-cooling (FC) mode, respectively. The cooling field and measuring field are both 1000Oe. The specific heat $C_p$ was measured using the Quantum Design Physical Properties Measurement System (PPMS) in the standard procedure.

The electric polarization $P$ was measured using the mPyro method with different $T_{end}$=2K-38K, respectively. Each sample was polished into a thin disk of 0.2mm in thickness and 10mm in in-plane dimension, and then sandwich-coated with Au layers as top and bottom electrodes, as shown in Fig.2(a). The measurement was performed using the Keithley 6514A and 6517 electrometers connected to the PPMS. In details, each sample was submitted to the PPMS and cooled down to ~100K. Then a poling electric field $E_{pole}$~10kV/cm was applied to the sample until the sample was down to $T_{end}$, at which the sample was then short-circuited for sufficient time (>30min) in order to release any charges accumulated on the sample surfaces or inside the sample. The recorded background current noise was less than 0.2pA. Then the sample was heated slowly at a warming rate up to a given temperature $T_0$=60K>$T_N$, during which the released current $I_{tot}$ was collected. Similar measurements were performed with different warming rates from 1K/min to 6K/min and the collected $I_{tot}$ data are compared to insure no contribution other than pyroelectric current $I_{pyro}$. Finally, polarization $P(T)$ was obtained by integrating the collected $I_{pyro}(T)$ data from $T_0$ down to $T_{end}$. The validity of this procedure was confirmed repeated in earlier works [15] and will be shown below too.

In addition, the dielectric susceptibility $\varepsilon$ at various frequencies as a function of $T$ was collected using the HP4294A impedance analyzer with an *ac*-bias field of ~50mV. Besides the $\varepsilon$-$T$ data and $P$-$T$ and data, we also measured the response of $P$ to magnetic field $H$ in two modes. One is the isothermal mode with which the variation in $P$ in response to the scanning of $H$ was detected and the other is the iso-field mode with which the $P$-$T$ data under a fixed $H$ were collected. By such measurements, one can evaluate the magnetoelectric (ME) coupling of the samples. We define $\Delta P(H)=P(H)-P(H=0)$ as the ME parameter.

We also employed the PUND method to obtain the P-E loops at various temperatures, using the identical procedure as reported in literature e.g. Ref.[45]. Our data are quite similar to reported ones from other groups. As an example, the evaluated $P$ at a pulsed field of 30kV/cm as a function of $T$ is presented in Fig.3(d).

## IV. Results and discussions

### A. Magnetic phase transitions

We first look at the phase transition sequence in terms of specific heat $C_P$, magnetization $M$, and dielectric susceptibility $\varepsilon$ as a function of $T$, as shown in Fig.5. For reference, the released current $I_{tot}(T)$ using the mPyro method at a warming rate of 2K/min is shown in Fig.5(d). The $C_P(T)$ curve shows clear anomalies roughly at $T_{N1}$~40K, $T_{N2}$~27K, and $T_{Dy}$~8K, while the peak at $T_{N3}$~20K, if any, is weak. These anomalies reflect the sequent magnetic transitions from the paramagnetic phase to the C-AFM phase, the coexisting IC-AFM phase plus C-AFM phase, the two IC-AFM coexisting phases, and the independent $Dy^{3+}$ spin ordering plus IC-AFM phase. These observations are consistent with earlier reports [36].

Surely, no features corresponding to these transitions, except the independent $Dy^{3+}$ spin ordering at $T_{Dy}$, can be observed in the measured $M$-$T$ data due to the fact that the $Dy^{3+}$ moment is much bigger than the $Mn^{3+}$/$Mn^{4+}$ moments and thus the signals are mainly from the $Dy^{3+}$ spins. The anomalies in the $\varepsilon$-$T$ curve at these phase transitions reflect the magneto-dielectric response, as revealed earlier [8]. Interestingly, a series of anomalies in the $I_{pyro}$-$T$ curve at these transitions are available, as shown in Fig.5(d), consistent with earlier report [8] too, evidencing the strong ME effect.

Comprehensive investigations revealed that these magnetic transitions arise from the competitions between complicate Mn-Mn, Dy-Mn, and Dy-Dy interactions [38]. The Mn-Mn interactions dominate the paramagnetic to C-AFM transitions at $T_{N1}$, and subsequently the Dy-Mn interactions become important and take part in determining the magnetic structures above $T_{Dy}$, below which the $Dy^{3+}$ spins order into the independent C-AFM state. Simultaneously, the IC-AFM Mn-spin structure above $T_{Dy}$ is modulated by this independent $Dy^{3+}$ ordering, resulting in a different IC-AFM order below $T_{Dy}$. These magnetic transitions may be of first-order, and the cooling run is different from the warming run, to be confirmed below again by our data on the electric polarization.

## B. Pyroelectric current anomalies

In Fig.6(a)~(c) are plotted the measured released currents from the sample using the mPyro method with $T_{end}$=2K, $E_{pole}$=10kV/cm, and three different warming rates, 2, 4, and 6K/min, respectively. It comes immediately to our attention that the three current-temperature curves, if normalized by the warming rate, almost perfectly overlap with each other, showing no difference between them within the measuring uncertainties of ~1.0pA and less than 0.5K peak-to-peak shift along the $T$-axis. These peaks are sharp and well fixed while thermally stimulated currents other than the pyroelectric current are usually broad. These features indicate that the measured data do come from the pyroelectric current $I_{pyro}$ without identifiable contribution from other sources. In addition, the measured $I_{pyro}$-$T$ curve can be switched upon a reverse poling field, as shown in Fig.6(d), indicating its origin from the pyroelectricity.

Moreover, the $I_{pyro}$-$T$ curves show clear anomalies at all the magnetic transition points ($T_{N1}$, $T_{N2}$, $T_{N3}$, and $T_{Dy}$), indicating the one-to-one correspondence between the magnetism and ferroelectricity as well as the ME response. These anomalies may be seriously smeared out upon the integration, which are thus hard to observe in the $P$-$T$ curves, suggesting that $I_{pyro}$ can be a parameter much more sensitive than $P$ at least in its response to phase transitions. The as-evaluated $P$-$T$ curves from the $I_{pyro}$-$T$ curves under $E_{pole}$=±10kV/cm in Fig.6(d) are plotted in Fig.6(e), and as expected, no clear anomalies at $T_{N2}$ and $T_{N3}$ in the $P$-$T$ curve can be observed. The $P$-$T$ curve at $E_{pole}$=10kV/cm is similar in shape to that reported in Ref.[8] although details of them may be different from each others, but completely different from that measured by the Pole and P-E methods [9] as well as by the PUND method. It is shown that the measured $P$ between $T_{Dy}$ and $T_{N1}$ is negative and the sign change occurs around $T_{Dy}$. Upon $T$ decreasing down to 2K from $T_{Dy}$, the $P$ tends to be saturated at 0.13mC/m$^2$, a sufficiently large polarization, suggesting that the X-phase is ferroelectric.

The sign change of $P$ from negative values to positive values with decreasing $T$ suggests immediately that DyMn$_2$O$_5$ is possibly a ferrielectric (FI) rather than a normal ferroelectric [8]. The sign change would be the consequence of competition between the two FE sublattices should DyMn$_2$O$_5$ be at a FI state. In this case, careful measurement using the mPyro method can provide critical data on details of the FI state as a function of $T$.

### C. Electric polarization

In proceeding, we present the measured $I_{pyro}$-$T$ curves for a series of $T_{end}$ but constant $E_{pole}$=10kV/cm in Fig.7, where the as-evaluated $P$-$T$ curves are also inserted. For reference, the $I_{pyro}$-$T$ curve with $T_{end}$=2K is shown by a thin dashed line in each plot, so that the differences of the $I_{pyro}$-$T$ curves from that with $T_{end}$=2K can be seen clearly.

It is observed that, given a constant $E_{pole}$, the $I_{pyro}$ and thus $P$ measured by this mPyro method are remarkably $T$-dependent. Upon increasing $T_{end}$ from 2K up to 15K, the $P$-$T$ curves all show the sign change, similar to the case with $T_{end}$=2K (see the left column of Fig.7). That $T_{end}$=12K implies that the sample was only cooled into the FE2 or FE3 phase instead of the X-phase. Both the FE2 and FE3 phases are ferroelectric with sufficiently large polarization, as demonstrated by all the Pyro [8], P-E [9], Pole [9], and PUND methods as well as mPyro method here.

The temperature $T_{P=0}$, at which the $P(T)$ changes its sign, is plotted as a function of $T_{end}$ in Fig.8(a). It is seen that $T_{P=0}$ has a shift of ~9K with a ~10K increasing in $T_{end}$, implying that the FI state, if applicable, is not robust against thermal fluctuations ($T$) or external field ($E_{pole}$). Supposing the FI state is composed of two or more sublattices, one expects that at least one of them is strongly $T$- or $E_{pole}$-dependent. What is surprising is the $I_{pyro}$-$T$ curves as $T_{end}$>12K, some of which are plotted in the right column of Fig.7. In spite of the positive poling field $E_{pole}$=10kV/cm, the measured $P$ is negative and no more sign change is observed. The negative peak at $T_{N1}$ remains nearly unchanged even with $T_{end}$=38K, very close to $T_{N1}$=40K. Such a negative polarization in a sample under a positive field poling can't be possible in a simple ferroelectric, unless the polarization has two or more components which are anti-parallel to each other, as shown in a ferrielectric state.

### D. Path-dependent behavior

Besides the results described above, it is observed that the $I_{pyro}$ and thus $P$ measured by this mPyro method are not only $T$-dependent but also $T_{end}$-dependent, or in the other words, path-dependent. This $T_{end}$-dependence seems to be strange at the first glance. It reflects the fact that the electric polarization has the magnetic origin, since no structural phase transitions occur below $T_{N1}$. One expects that this path-dependence may be related to the first-order magnetic transitions and thus the magnetic structure evolution is path-dependent. To illustrate this path-dependence, we present in Fig.8(b) the measured $P$ value at $T_{end}$, i.e. $P(T_{end})$, where the $P$ value as a function of $T$ with $T_{end}$=2K, i.e. $P(T)$, is inserted for comparison by setting $T=T_{end}$. For a normal ferroelectric, $P(T_{end})$ should overlap with $P(T)$ for $T_{end}=T$. Here, it is clearly shown that $P(T_{end})$ coincides with $P(T)$ only below $T_{Dy}$ and they separate from each other at $T_{Dy}$ and above, noting that $P(T_{end})$ is always above $P(T)$. The difference between them maximizes at $T_{end}=T\sim$10K and ~24K, while becomes negligible as $T_{end}\rightarrow T_{N1}$, suggesting that the magnetic transitions below $T_{N1}$ have the first-order characteristic. In fact, combining the $P(T_{end})$ and $P(T)$ data generates a double-loop like (not typical) hysteresis, as shown in Fig.8(b), which can be a characteristic of such first-order phase transitions in the sense of ME coupling.

All the results presented above allow us to argue that for $DyMn_2O_5$, besides the ferrielectric state with more than one ferroelectric component, the electric polarization shows remarkable path-dependence. These phenomena provide additional evidences with the argument that the physics of multiferroicity not only in $DyMn_2O_5$ is complicated. We propose a FI model in next section to interpret the observed phenomena.

### V. Ferrielectric state

#### A. Two FE sublattices

It has been well believed that for $DyMn_2O_5$, the symmetric exchange striction effect arising from the specific AFM Mn-Mn and Dy-Mn spin alignments, is the main mechanism for generating the electric polarization, while the asymmetric exchange striction arising from the spiral Mn spin order was reported to be non-negligible [16]. Although three subsequent transitions between the C-AFM orders and IC-AFM orders occur upon varying $T$, the spin configurations in these spin ordered phases don't show big difference, and both the C-AFM and IC-AFM phases have been shown to be ferroelectric. For simplification consideration, we don't take into account in this work the contributions from the spiral spin order and C-AFM/IC-AFM transitions, leaving them for future clarifications. However, the effect of the independent Dy spin ordering at $T_{Dy}$ imposes significant effect on the electric polarization, due to the strong Dy-Mn interactions, which will be considered.

Referring to relevant literature on $DyMn_2O_5$ [34], we present in Fig.9(a) a schematic drawing of the spin structure projected on the *ab*-plane, which varies a little over the *T*-range between $T_{N1}$ and $T_{Dy}$. The $Dy^{3+}$, $Mn^{3+}$, and $Mn^{4+}$ ions and spins are denoted using different color dots and arrows. The square pyramid and octahedral structural units surrounding these spins are drawn too for a better view, as sampled in Fig.9(b). It is noted that the light gray and gray structural units are located on two different *ab*-planes which shift 1/4 lattice unit from each other along the *c*-axis [16]. If looking at the geometric structure along the *b*-axis, one finds two types of three-spin blocks each centered on a $Mn^{4+}$ spin in the octahedral unit, as shown in Fig.9(c) and (d), respectively. One is the block A, consisting of one $Mn^{4+}$-centered octahedral connected with two pyramid units each with one $Mn^{3+}$ spin inside (Fig.9(c)). The other is the block B, consisting of one $Mn^{4+}$-centered octahedral connected with two $Dy^{3+}$ spins located in the space surrounded by the $MnO_6$ and $MnO_5$ units (Fig.9(d)). Because of the symmetric exchange striction effect, the two $Mn^{3+}$ ions in block A shift roughly up and the two $Dy^{3+}$ ions in block B shift down with respect to the $Mn^{4+}$ ions. Therefore, one electric polarization component ($P_{MM}$) in block A and one polarization component ($P_{DM}$) in block B are generated. They are anti-parallel to each other but roughly align along the *b*-axis. Here we don't discuss the possible polarization along the *a*-axis. The whole lattice as the consequence of the alternating stacking of the two types of blocks is therefore ferrielectric composed two FE sublattices.

It should be noted that different from $P_{MM}$, $P_{DM}$ originates from the Dy-Mn interactions and thus depends on the $Dy^{3+}$ spin ordering. Besides the independent spin ordering below $T_{Dy}$, $Dy^{3+}$ spins may order in coherence with the Mn spin ordering due to the strong Dy-Mn interactions. This $Dy^{3+}$ spin ordering occurs roughly around $T_{N1}$ or $T_{N2}$, leading to the spin configuration shown in Fig.9(a). However, this induced $Dy^{3+}$ spin order may be partially and gradually replaced by the independent ordering below $T_{Dy}$, details of which have not been well understood so far. If this scenario applies, one can expect that $P_{DM}$ will show much more significant *T*-dependence than $P_{MM}$, since the $Dy^{3+}$ spin ordering above $T_{Dy}$ is induced by the Mn spin order. Also, one may see the modulation of $P_{DM}$ upon the entrance of independent $Dy^{3+}$ spin ordering below $T_{Dy}$, while the Mn spin configuration may be slightly affected by this $Dy^{3+}$ spin ordering too.

## B. Temperature dependences of $P_{DM}$ and $P_{MM}$

The above discussion stimulates us to check the $P_{DM}$ and $P_{MM}$ as a function of $T$, respectively. To evaluate these dependences, one again consults to the $P$-$T$ data shown in Fig.7. For simplification, the effect of independent $Dy^{3+}$ spin ordering below $T_{Dy}$ on the Mn spin order is assumed to be weak if any. In this case, it can be reasonably assumed that $P_{MM}$ initiating at $T_{N1}$ increases rapidly with decreasing $T$ and becomes saturated in the low $T$ range, because the Mn spin order is already well developed right below $T_{N1}$. Consequently, $P_{DM}$ as a function of $T$ can be extracted. Take the data with $T_{end}$=8K$\geq T_{Dy}$ as an example. The measured $P(T)$ data are plotted in Fig.10(a). The $P_{MM}(T)$ curve is extracted based on the above assumption, and then $P_{DM}(T)=P-P_{MM}$ is evaluated. For clear illustration, the two ferroelectric sublattices on the $ab$-plane are schematically drawn in Fig.10(c) and (d), and an overlap of them constitutes the ferrielectric lattice in Fig.10(b). As expected, it is seen that $P_{DM}$ increases slowly with decreasing $T$ until $T\sim$20K and then rapidly, exhibiting much more significant $T$-dependence than $P_{MM}$.

We again take the data with $T_{end}$=2K$<<T_{Dy}$ as an example to illustrate the effect of the independent $Dy^{3+}$ spin ordering on the $P_{DM}$. The results are shown in Fig.11. In this case, at $T=T_{end}$, partial $Dy^{3+}$ spins are on the track of the independent spin order, melting away local polarizations at some lattice sites, as shown in Fig.11(b). The FE sublattice $P_{DM}$ is thus partially melted away, giving rise to a smaller $P_{DM}$. This is the reason for the lower $T_{P=0}$ and smaller $|P|$ below $T_{P=0}$ with respect to the case of $T_{end}$=8K.

Here it should be mentioned that the difference in the $P_{DM}(T)$ curve between the case of $T_{end}$=8K and that of $T_{end}$=2K reflects the difference in the magnetic structures for the two cases. This is different from the observations in normal ferroelectrics where no such difference should be available. The origin lies in the fact that the magnetic transitions at $T_{Dy}$ are the first-order. Due to lack of details of this first-order magnetic transition, no full understanding of this path dependence is available at this stage, which deserves for future investigation.

*C. Discussion on negative $P_{MM}$*

Given the ferrielectric model shown in Fig.9 and the different temperature dependences of $P_{DM}$ and $P_{MM}$ shown in Fig.10 and Fig.11, a puzzling issue appears, that the measured $P_{MM}$ (or $P$) remains to be negative even so $T_{end}$ is higher than $T_{N2}$. It is reasonably expected that $P_{DM}$ should be much smaller than $P_{MM}$ and thus a poling by a positive $E_{pole}$ would allow the up-alignment of $P_{MM}$, opposite to the case shown in Fig.10(b). In this case, the measured pyroelectric current $I_{pyro}$ and polarization $P$ should be positive, while the measured data contradict with this prediction.

For illustration, one presents the data for $T_{end}$=25K in Fig.12, while the measured $P(T)$ is negative. By the identical procedure, one evaluates respectively the $P_{MM}(T)$ and $P_{DM}(T)$ (blue and olive lines), noting that the pyroelectric current is measured under zero electric bias ($E$=0). It is seen that $|P_{MM}|>|P_{DM}|$ over the whole-$T$ range. It implies that a positive electric poling can't align the $P_{MM}$ ferroelectric sublattice along the direction of $E_{pole}$, which is strange and anomalous.

At this stage, we have no convincing explanation of this anomalous phenomenon. One possible reason is that $P_{DM}$ as a function of $T$ is very sensitive to poling field $E_{pole}$. Considering the fact that the $Dy^{3+}$ spins are much softer (or elastic) than the Mn spins in response to magnetic field, one expects that the electric field driven alignment of $Dy^{3+}$ spins coherently with the Mn spins into the configuration shown in Fig.9(a) would be energetically easy. Therefore, $P_{DM}$ can be remarkably enhanced by $E_{pole}$. If it is the case, the electric poling during the cooling sequence enhances $P_{DM}$ remarkably while $P_{MM}$ is roughly unchanged, so that $P_{DM}$ around $T_{end}$ is larger than $P_{MM}$ in magnitude, resulting in the alignment of $P_{MM}$ opposite to $E_{pole}$ and $P_{DM}$. After the removal of $E_{pole}$ at $T_{end}$, $P_{DM}$ shrinks back to the value at $E$=0, much smaller than $P_{MM}$. Consequently, the pyroelectric current remains negative.

Another possible explanation for this strange phenomenon is the ferroelastic effect of $DyMn_2O_5$ [46], which results in clamping the $P_{MM}$ ferroelectric domains along the opposite direction to $E_{pole}$ during the poling process. However, this effect remains to be an issue and no evidence as well as its details are available.

*D. PUND measurement*

For providing additional evidence with the ferrielectricity at low $T$, we employ the PUND method to measure the $P$-$E$ loops at different $T$. Experimentally, no ideal double-loop hysteresis even for a typical ferrielectric with well defined and non-interactive two FE sublattices can be often observed. In most cases, the double-loop feature is seriously smeared out, and usually a deformed singe-loop hysteresis is expected, in particular at extremely low $T$ at which the coercive field is high and the domains are easily trapped. In the present work, the samples are polycrystalline and thus the measured $P$-$E$ data is more like a single-loop hysteresis, unfortunately.

The measurement of the $P$-$E$ loops was based on our careful calibration of the measuring unit following the standard procedure [45]. While no systematic presentation of all the data is intended here, we show two $P$-$E$ loops obtained at two different $T$ in Fig.13(a) and (b). It is seen that the loop is anomaly deformed, different from that of a normal ferroelectric. The coercive field is bigger than the reported value in Ref.[9] for single crystals, while it should be smaller. An extremely big coercive field is usually true for a ferrielectric.

To illustrate the features of this observed $P$-$E$ hysteresis, we start from an ideal double-loop for a ferrielectric, as shown in Fig.13(c). The evolution of this double-loop hysteresis upon decreasing $T$ or from single crystal sample to polycrystalline one is indicated by the arrows. It is seen that the measured $P$-$E$ loop is quite similar to the hysteresis observed here.

*E. Magnetoelectric response*

Obviously, referring to the ferrielectric model shown in Fig.9, one immediately predicts that the ME parameter $\Delta P$ is negative and remarkably $T$-dependent. It is well known that the 4$f$ spins in most transition metal oxides is sensitive to magnetic field, and even at extremely low $T$ a field as big as ~2.0T is sufficient to align the 4$f$-spins [9]. For DyMn$_2$O$_5$ here, this suggests that a low magnetic field is sufficient to re-align the Dy$^{3+}$ spins along the field direction, while the Mn spins remain highly robust against magnetic field. Therefore, the ↑↑↓ or ↓↓↑ pattern in block B is melt out, as shown in Fig.14, leading to disappearance of $P_{DM}$ and roughly unchanged $P_{MM}$.

The experimental data conform well this scenario, as presented in Fig.15. Fig.15(a) plots the $P$-$T$ data at $T_{end}$=2K, where components $P_{DM}(T)$ and $P_{MM}(T)$ under $H$=0 are presented too. The $\Delta P(H>2T)$ should not be much less than but almost equivalent to $P_{DM}$ in magnitude although their signs are opposite. Our data do support this prediction even in the quantitatively sense, as shown in Fig.15(b), (c), and (d), respectively. The ferrielectric state as a basis for this ME effect is then accommodated.

*F. Remarks*

To this stage, we have presented a qualitative explanation of the major features associated with the electric polarization and ME effect observed in DyMn$_2$O$_5$ based on the ferrielectricity model as concretized in this work. Nevertheless, due to either lacking of sufficient data or uncertainties in details of this model and our understanding, several issues remain unclear or unsolved:

(1) No detailed discussion on the possible ferroelectric phase transitions at the magnetic transition points $T_{N2}$, $T_{N3}$, and even $T_{Dy}$, respectively, has been given. The magnetic structure shown in Fig.9 is more or less a qualitative description of the state at $T_{Dy}<T<T_{N2}$. It is clear that the magnetic structure is a little different in various $T$-ranges, which should be mapped into the weak anomalies at these transition points, as reflected in the $I_{pyro}$-$T$ curves and then can be reasonably described upon sufficient data on the magnetic structure details in these $T$-ranges. Also, the path-dependence of the electric polarization is attributed to the magnetic transitions which should be path-dependent due to the first order characteristic. This characteristic remains to be clarified.

(2) An uncertain point regarding the present ferrielectric model is the response of $P_{DM}$ to electric field which is assumed to be remarkable in order to account for the experimental observations. Searching for convincing evidence with this assumption is challenging although the assumption itself is physically reasonable.

(3) Besides the symmetric exchange striction mechanisms, DyMn$_2$O$_5$ also accommodates the asymmetric exchange striction mechanism [16], which can be seen in Fig.9(a) where the spin alignment in the Mn$^{3+}$-Mn$^{3+}$-Mn$^{4+}$-Mn$^{3+}$-Mn$^{3+}$-Mn$^{4+}$-…chains along the *b*-axis is spiral. This spiral spin order allows possibly a local polarization along the *a*-axis. However, it seems that the polarizations between the two neighboring chains cancel with each other. This issue surely deserves for further attentions. In this sense, the present ferrielectric model does not account of the contribution from the spiral spin order.

(4) A thermodynamic description in the phenomenological sense for the ferroelectricity in RMn$_2$O$_5$ family was already proposed in Ref.[46]. A consistent description accounting for the ferrielectric state certainly deserves for further investigation. At the same time, the first-principles calculation on this specific ferrielectric state is also necessary to confirm the existence of ferrielectricity.

(5) The existence of a ferrielectric state in DyMn$_2$O$_5$ may not be a specific case for the RMn$_2$O$_5$ family and it may also apply to other members with the magnetic R ion, or even with non-magnetic ion too. However, due to the competitions among various interactions, the structural model shown in Fig.9 may not be of generality. It is expected that the electric polarization in these multiferroics exhibits more complicated behaviors than normal ferroelectrics, and any direct extension may be cautious.

## VI. Conclusion

In conclusion, extensive pyroelectric current measurements on DyMn$_2$O$_5$ as a function of temperature and magnetic field have been performed, and the complicated polarization behaviors have been characterized. It is revealed that the electric polarization in DyMn$_2$O$_5$ does consist of two antiparallel components, demonstrating the ferrielectric state instead of a ferroelectric state at low temperature. The two polarization components are believed to originate from the symmetric exchange striction mechanism. One is generated from the Mn$^{3+}$-Mn$^{4+}$-Mn$^{3+}$ blocks with the ↓↑↑ and ↑↓↓ spin alignments, which is robust against temperature and magnetic field. The other is generated from the Dy$^{3+}$-Mn$^{3+}$-Dy$^{3+}$ blocks with the ↓↓↑ and ↑↑↓ spin alignments, which is remarkably temperature-dependent and sensitive to magnetic field. The observed magnetoelectric effect is mainly attributed to the spin re-alignment of the Dy$^{3+}$ spins in response to magnetic field, resulting in the partial melting of the ↓↓↑ and ↑↑↓ spin alignments in the Dy$^{3+}$-Mn$^{3+}$-Dy$^{3+}$ blocks. The present work represents a substantial step towards a full-scale understanding of the electric polarization in DyMn$_2$O$_5$ and other RMn$_2$O$_5$ family members too.

**Acknowledgement**:

This work was supported by the National 973 Projects of China (Grants No. 2011CB922101 and No. 2009CB623303), the Natural Science Foundation of China (Grants No. 11234005 and No. 11074113), and the Priority Academic Program Development of Jiangsu Higher Education Institutions, China.

*Figure Captions*

Figure 1 (color online). Schematic drawing of the lattice structure of DyMn2O5 with the three major Mn-Mn spin interactions $J_3$, $J_4$, and $J_5$. The ions and coordinates are labeled for guide of eyes.

Figure 2 (color online). Schematic illustrations of the two methods used for probing electric polarization. (a) The pyro method in which the released current $I_{tot}$ of the capacitor as a function of temperature $T$ during the warming sequence is probed after the capacitor is cooled down to $T_{end}$ under electric poling by *dc* field $E_{pole}$. (b) The Pole method in which the current $I_{tot}$ flowing across the capacitor under a relatively low *dc* field $E_{pole}$ is measured during the cooling sequence. Here, $I_{tot}$ includes the leaky current $I_{leaky}$ and pyroelectric current $I_{pyro}$.

Figure 3 (color online). Evaluated electric polarizations measured by various methods, as a function of $T$, respectively. (a) The $P_{pyro}$-$T$ curve by the Pyro method, taken from Ref.[8], (b) the $P_{P\text{-}E}$-$T$ curve by the *P-E* hysteresis method, taken from Ref.[9], (c) the $P_{pole}$-$T$ curve by the Pole method, taken from Ref.[9], and (d) the $P_{PUND}$-$T$ curve by the PUND method, described in the text. Here $T_{N1}$, $T_{N2}$, $T_{N3}$, and $T_{Dy}$ are the magnetic transition points, and $T_{FE1}$, $T_{FE2}$, and $T_{FE3}$, are the ferroelectric transition points, see the text. Symbol PE refers to the paraelectrics phase. Symbols FE1, FE2, and FE3, refer to the three ferroelectric phases, respectively. The X-phase, as defined in Ref.[9], refers to the claimed low-$T$ non-ferroelectric phase.

Figure 4 (color online). Measured θ-2θ XRD spectrum for the DyMn$_2$O$_5$ sample. The calculated one by the Rietveld analysis and the Bragg positions of the reflections are inserted for reference. The evaluated lattice constants are given in the figure too.

Figure 5 (color online). Measured specific heat normalized by temperature ($C_P/T$), magnetizations (*M*) under the ZFC and FC conditions, dielectric constant ($\varepsilon$), and released current ($I_{tot}$) by the Pyro method at a warming rate of 2K/min, are presented in (a), (b), (c), and (d), respectively. It is noted that the dielectric constant was measured at frequency of 100kHz, with no remarkable frequency dispersion.

Figure 6 (color online). Measured pyroelectric current $I_{pyro}$ as a function of $T$ at a warming rate of 2K/min (a), 4K/min (b), and 6K/min (c), respectively. The $I_{pyro}$-$T$ curves measured under two opposite poling fields ±10kV/cm at 2K/min warming rate as well as the evaluated polarizations $P_{pyro}(T)$ are plotted in (d) and (e) respectively.

Figure 7 (color online). Measured pyroelectric current $I_{pyro}=I_P$ and evaluated polarization $P$ as a function of $T$ at $T_{end}$=2K (a), 6K (b), 8K (c), 10K (d), 15K (e), 25K (f), 33K (g), and 38K (h), respectively. The warming rate is 2K/min. For reference, the $I_P$-$T$ data at $T_{end}$=2K are inserted.

Figure 8 (color online). (a) Evaluated crossing temperature $T_{P=0}$ at which the measured $P(T)$ changes its sign, as a function of $T_{end}$. (b) Evaluated $P(T)$ curve and $P(T_{end})$ curve. The warming rate for the pyroelectric current probing is 2K/min.

Figure 9 (color online). Proposed spin structure at a temperature lower than $T_{N2}$ and higher than $T_{Dy}$, referring to neutron scattering data available in literature. (a) The spin structure projected on the *ab*-plane with the square pyramid $Mn^{3+}$-$O^{2-}$ unit and octahedral $Mn^{4+}$-$O^{2-}$ unit shown in (b). The structural block A, composed of one $Mn^{4+}$-$O^{2-}$ octahedral connected by two $Mn^{3+}$-$O^{2-}$ pyramids roughly along the *b*-axis, is shown in (c). The structural block B, composed of one $Mn^{4+}$-$O^{2-}$ octahedral connected by two $Dy^{3+}$ roughly along the *b*-axis, is shown in (d). The proposed polarizations $P_{MM}$ and $P_{DM}$ generated by the two types of blocks due to the symmetric exchange strictions, are labeled in (c) and (d), respectively.

Figure 10 (color online). (a) Evaluated electric polarizations $P_{MM}$ and $P_{DM}$ from the two ferroelectric sublattices of the proposed ferrielectric lattice, as a function of $T$, where $P=P_{DM}+P_{MM}$ and $T_{end}=8K$. The proposed ferrielectric lattice and the associated two sublattices, all projected on the *ab*-plane, are schematically drawn in (b), (c), and (d), respectively.

Figure 11 (color online). (a) Evaluated electric polarizations $P_{MM}$ and $P_{DM}$ from the two ferroelectric sublattices of the proposed ferrielectric lattice, as a function of $T$, where $P=P_{DM}+P_{MM}$ and $T_{end}=2K$. The proposed two sublattices, projected on the *ab*-plane, are schematically drawn in (b) and (c), respectively.

Figure 12 (color online). Proposed $P_{DM}$ as a function of $T$ under zero electric field ($P_{DM}$ at $E=0$) and non-zero electric field ($P_{DM}$ at $E>0$). It is suggested that $P_{DM}$ is sensitive to external electric field $E$, implying that it has a much larger value during the cooling process with electric field poling than that during the warming process with no electric bias. The $P(T)$ and $P_{MM}(T)$ curves are inserted for reference. $T_{end}=25K$.

Figure 13 (color online). Measured *P-E* loops at $T=28K$ (a) and $T=13K$ (b), by the PUND method. The predicted *P-E* hysteresis for a ferrielectric lattice in single crystal or high $T$ (left, double loop) and polycrystalline sample or low $T$ (right, single deformed loop), with an intermediate state of the hysteresis, is shown in (c).

Figure 14 (color online). Schematic drawing of the spin alignments in block A and block B, respectively, under a downward magnetic field $H$ at $T<T_{Dy}$, referring to Fig.9. The $Dy^{3+}$ spins can be easily re-aligned by $H$, while the Mn spins can't, implying that $P_{DM}=0$ at $T<T_{Dy}$.

Figure 15 (color online). Measured magnetoelectric responses and proposed model. (a) The measured $P(T)$ curves and proposed $P_{MM}(T)$ and $P_{DM}(T)$ curves under $H=0$ and $H>>0$ (e.g. ~2T). It is suggested that $P_{MM}$ is robust against $H$ while $P_{DM}$ can be seriously suppressed by $H$, due to the field induced $Dy^{3+}$ spin realignment as proposed in Fig.14. The ferrielectric lattice at $H=0$ is shown in (b), which transfers into the lattice in (c) at $H>>0$. This lattice in (c) is composed of the $P_{MM}$ sublattice shown in (d) plus $P_{DM}$ sublattice shown in (e). $P=P_{DM}+P_{MM}$.

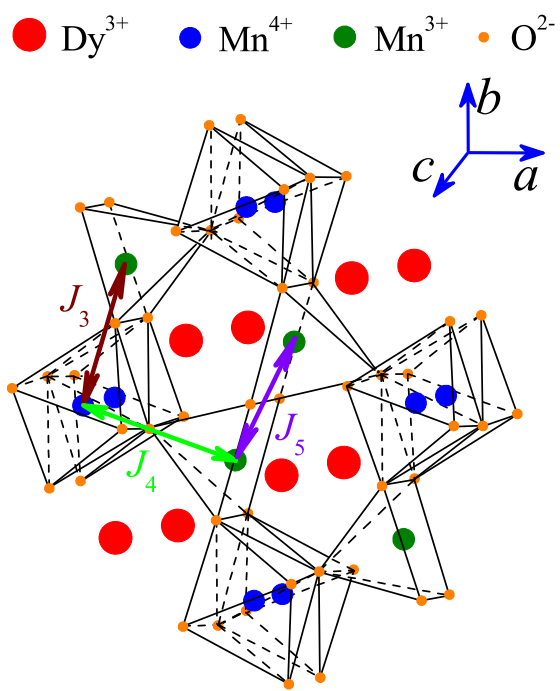

Figure 2

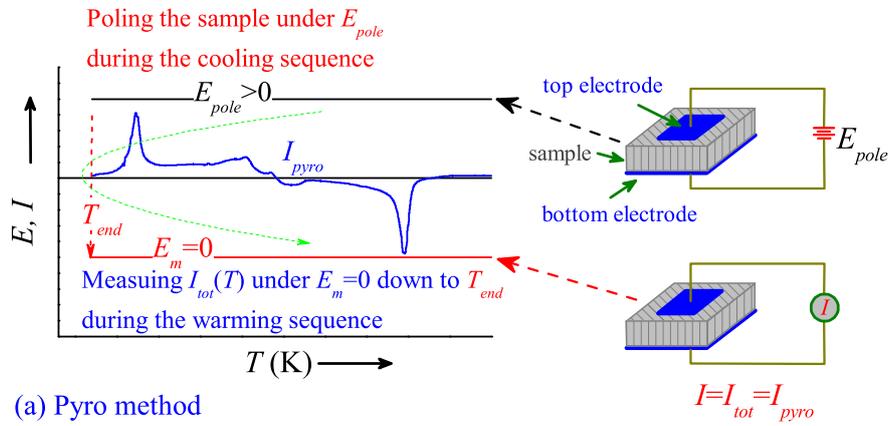

(a) Pyro method

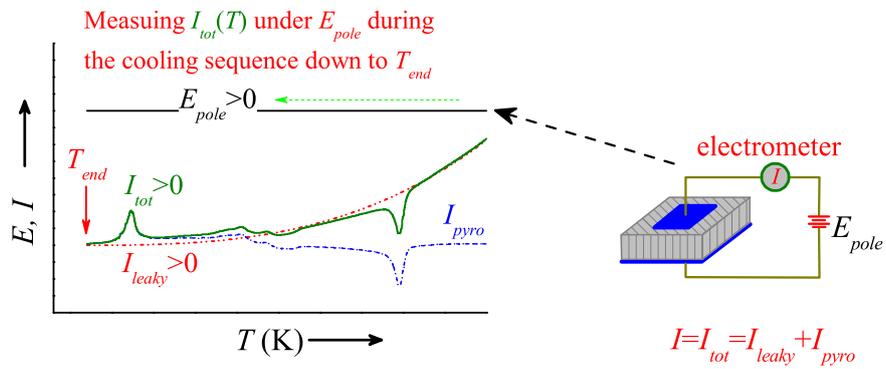

(b) Pole method

Figure 3

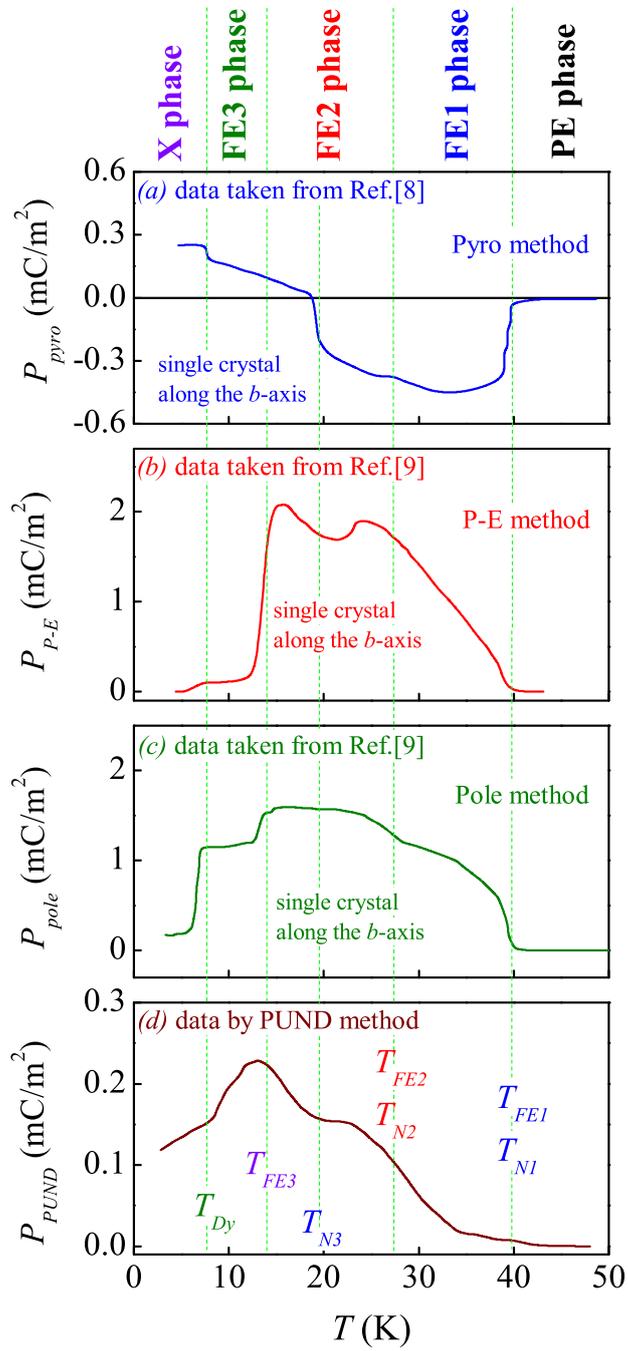

Figure 4

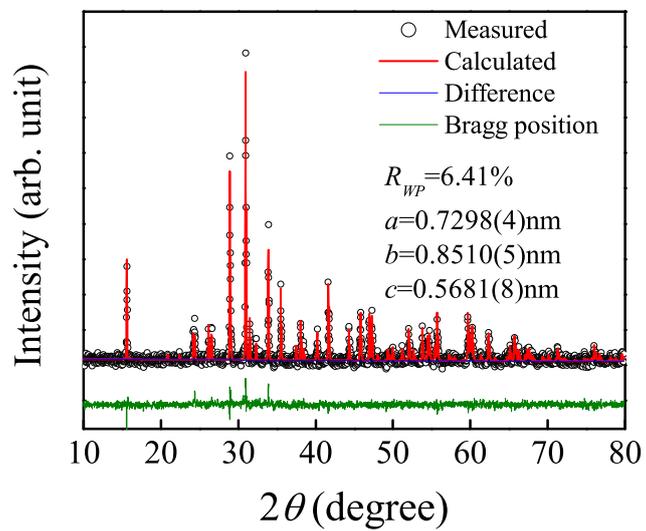

Figure 5

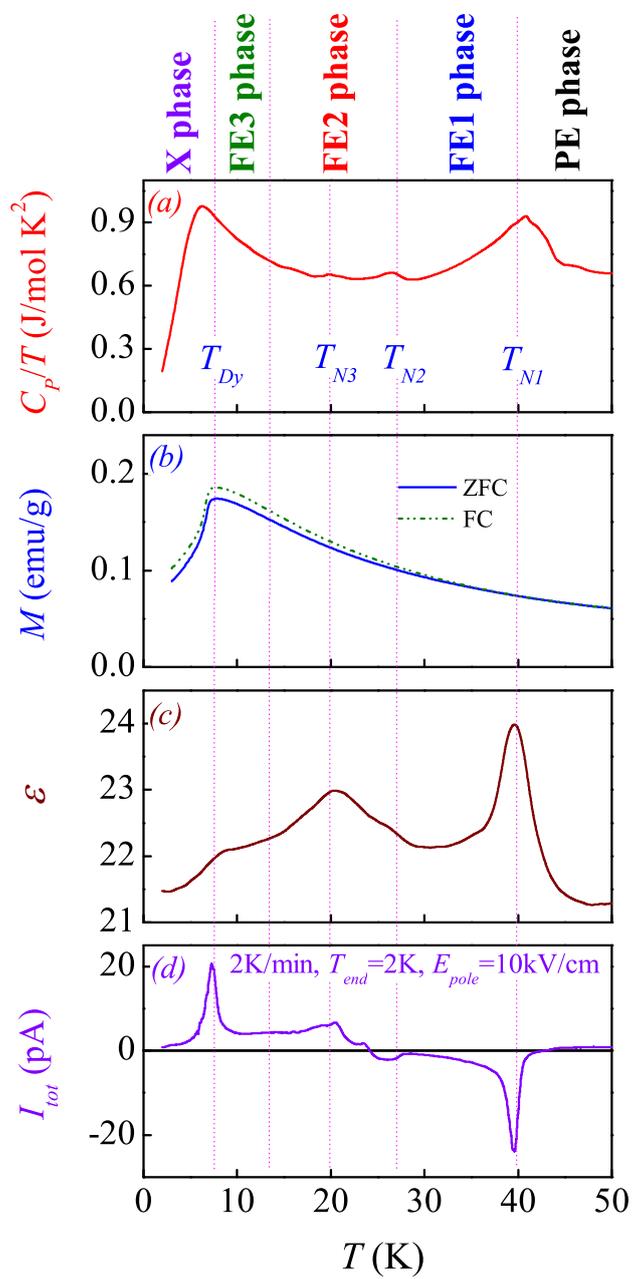

Figure 6

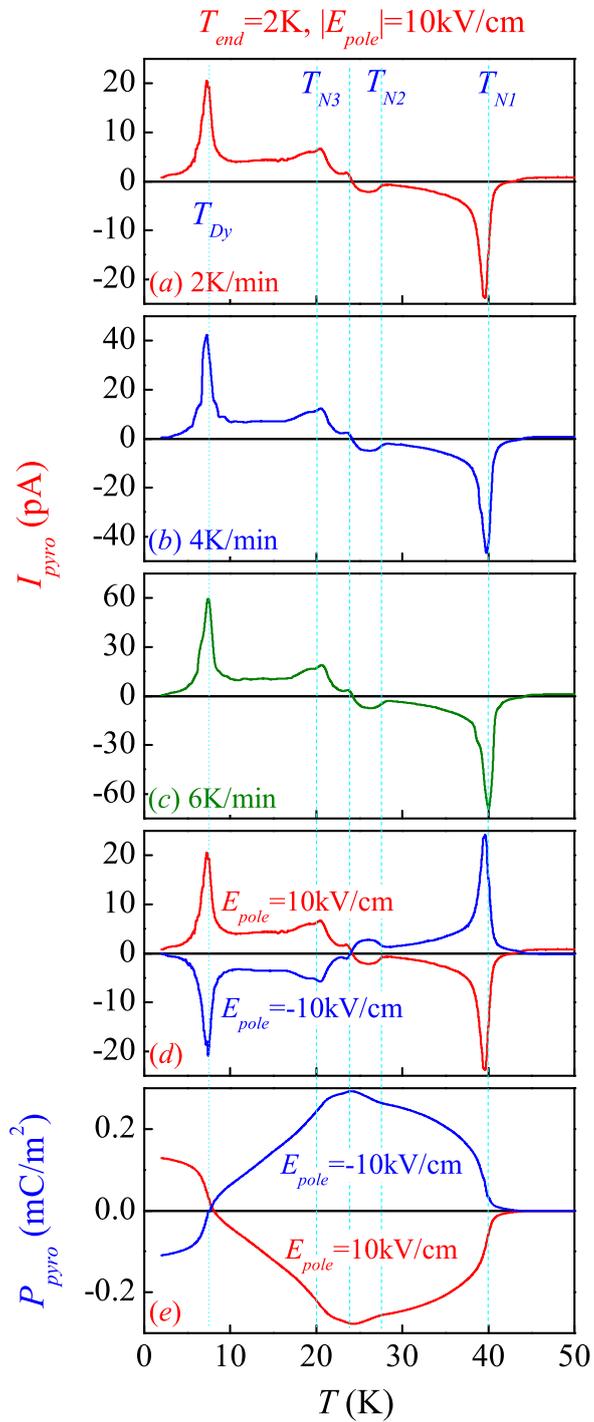

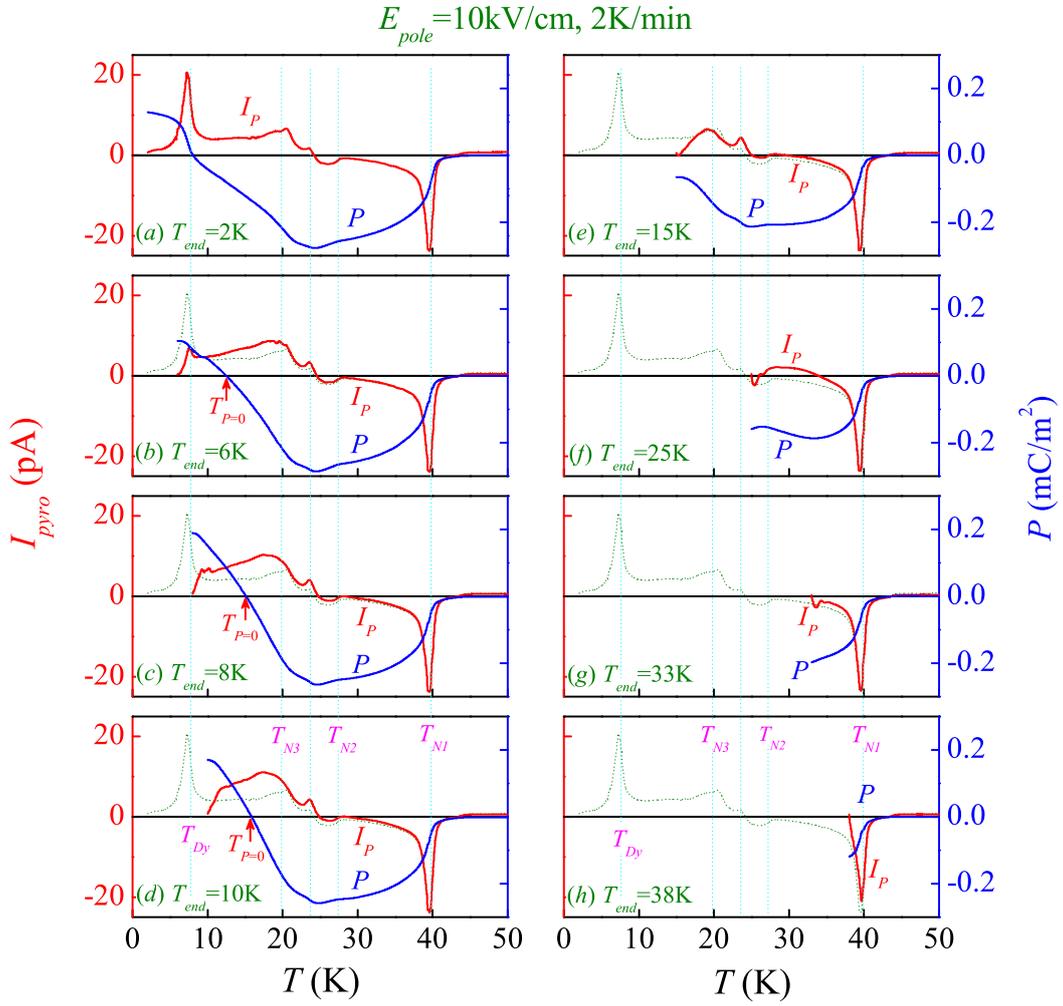

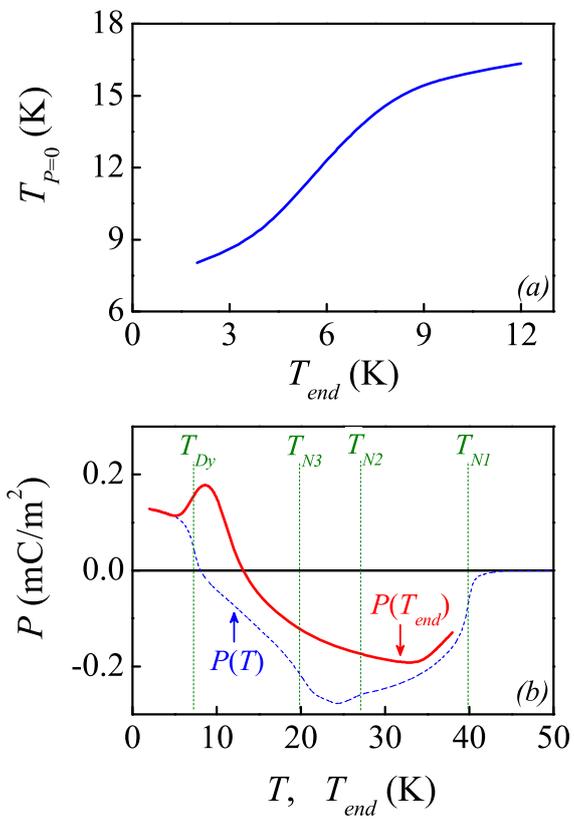

Figure 9

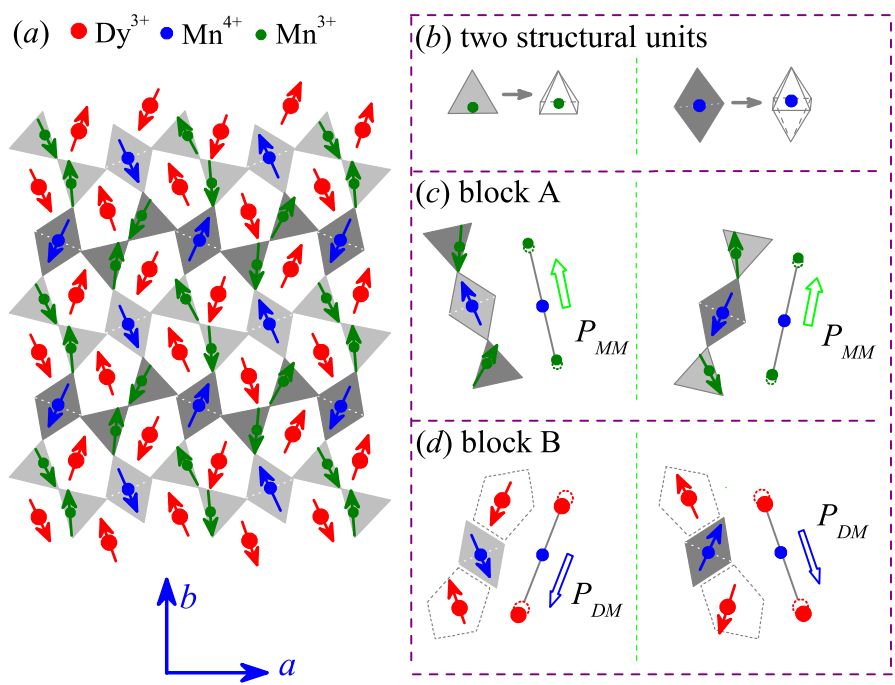

Figure 10

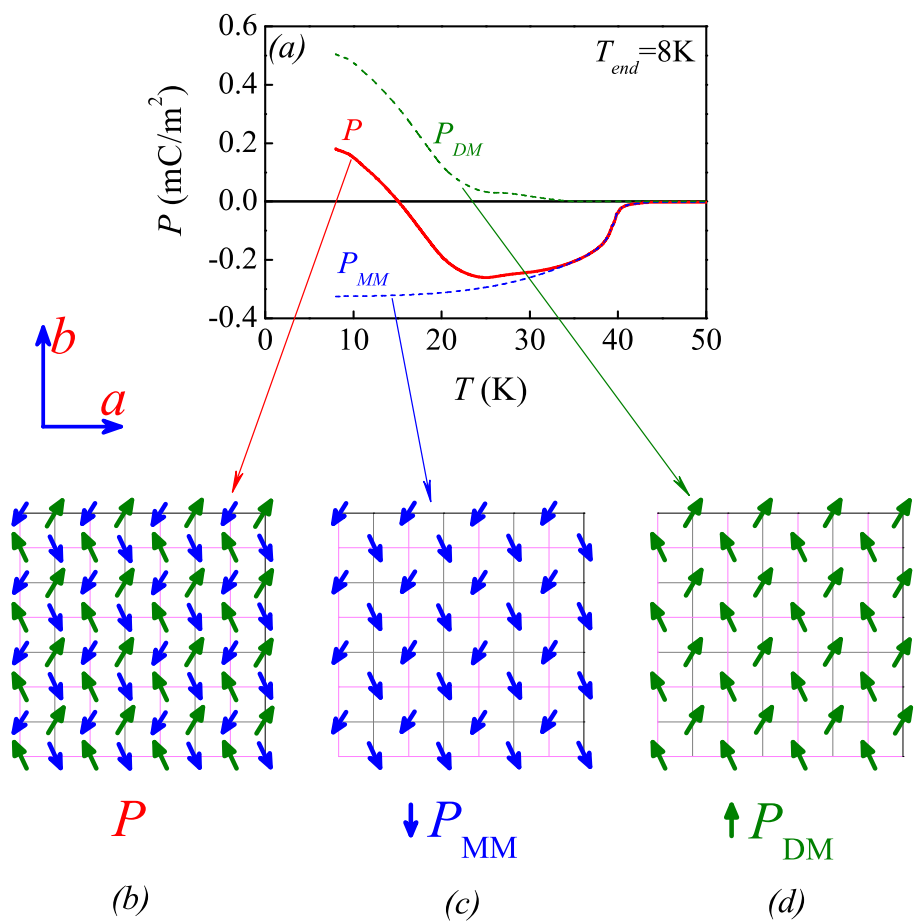

(b) P  (c) ↓P<sub>MM</sub>  (d) ↑P<sub>DM</sub>

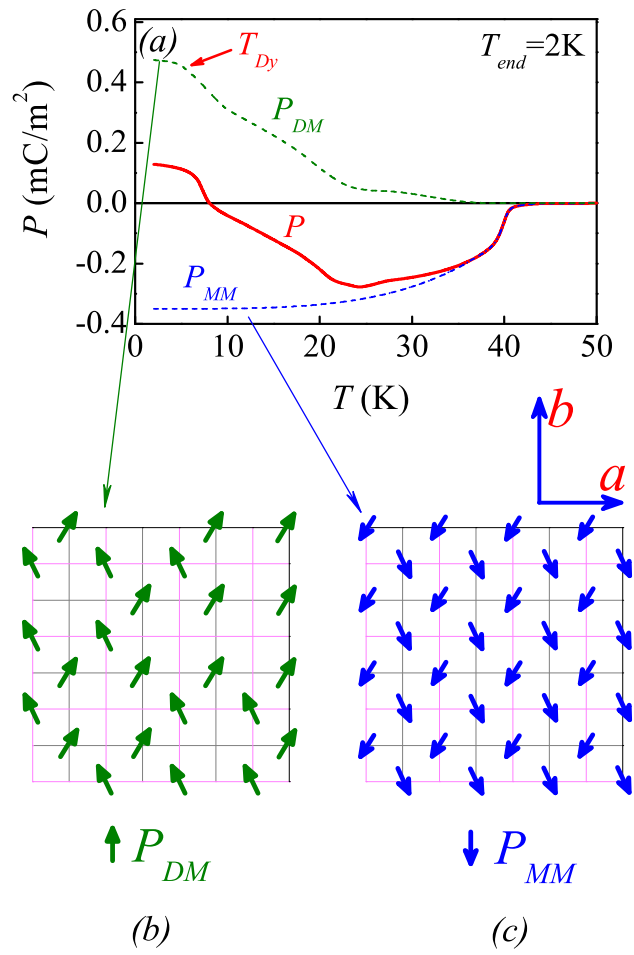

(b)                      (c)

Figure 12

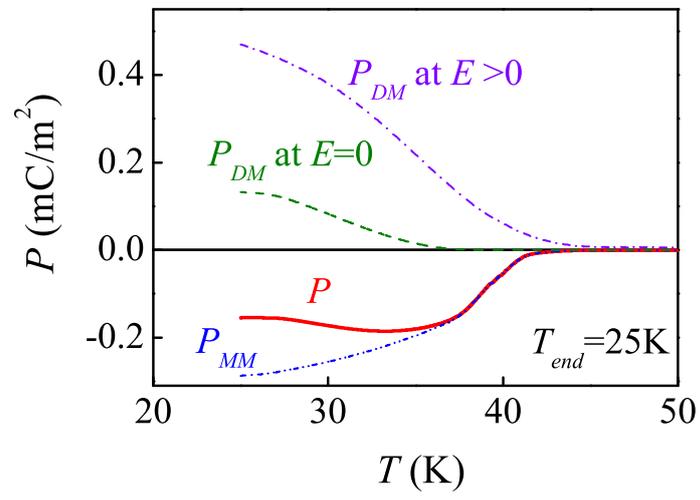

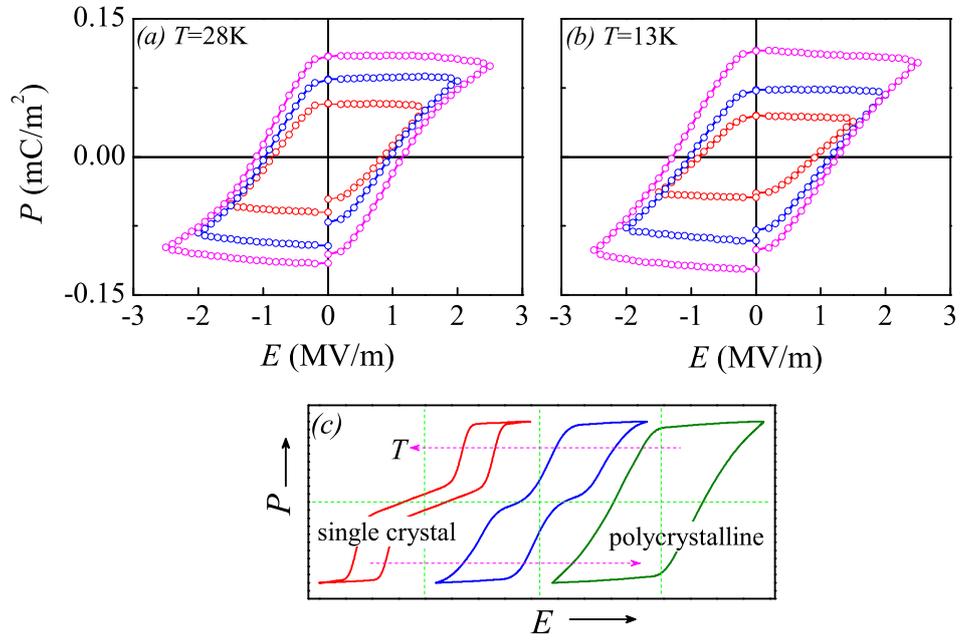

Figure 14

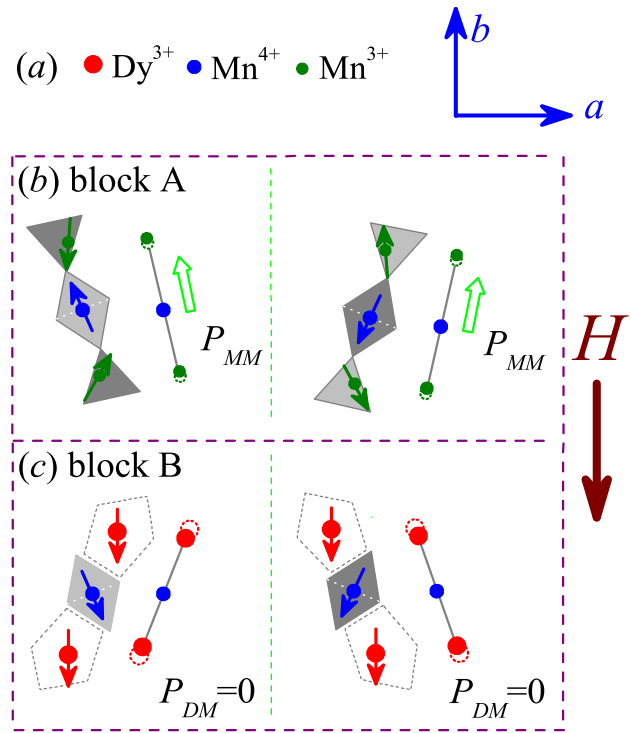

Figure 15

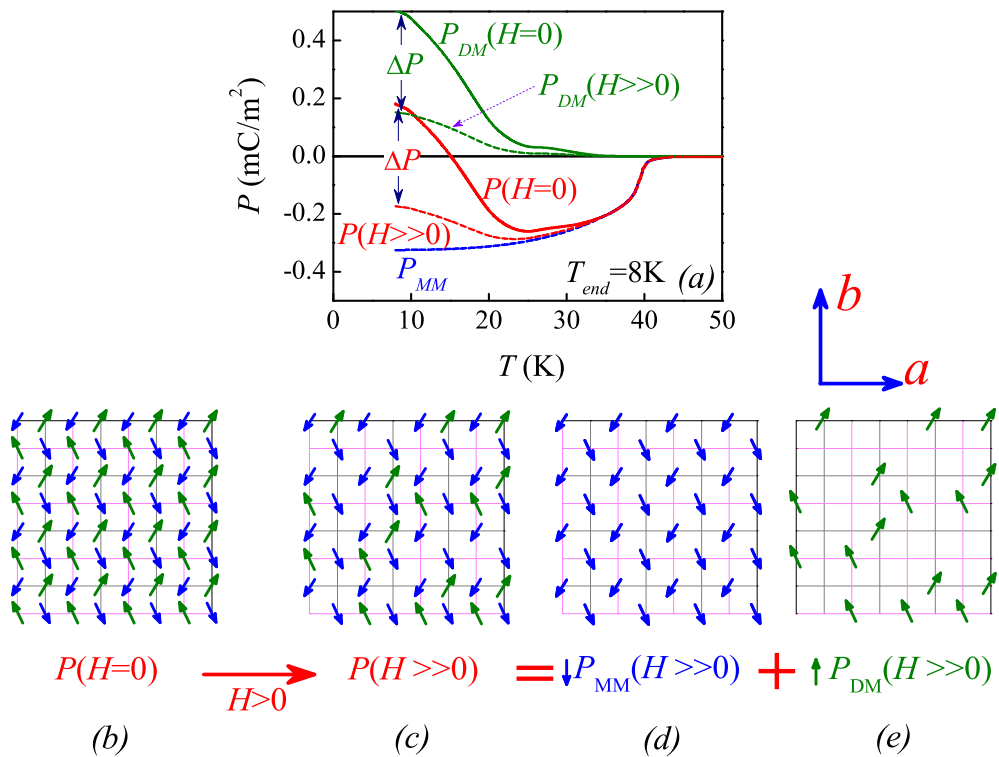